\documentclass{icst}

\usepackage[english]{babel}
\usepackage[utf8]{inputenc}
\usepackage{amsmath}
\usepackage{graphicx}
\usepackage[colorinlistoftodos]{todonotes}
\usepackage{multirow}

\usepackage[breaklinks,colorlinks,bookmarksopen,bookmarksnumbered,linkcolor=ICSTblue,citecolor=ICSTblue,urlcolor=ICSTblue]{hyperref}

\newcommand\BibTeX{{\rmfamily B\kern-.05em \textsc{i\kern-.025em b}\kern-.08em
T\kern-.1667em\lower.7ex\hbox{E}\kern-.125emX}}

\def\volumeyear{2018}

\journalname{XXXXXX}
\articletype{Research Article}
 \def\volumeyear{XXXX}
 \def\volumenumber{XX}
  \def\issuemonth{XX-XX}
  \def\issuenumber{X}
  \def\articleid{XX}
  \def\copyrightholder{L.A. Maglaras\textit{ et al.}}
\def\articledoi{10.4108/XX.X.X.XX}
\received{XXXX}
  \accepted{XXXX}
  \published{XXXX}
\begin{document}
\runningheads{L. A. Maglaras et al}{Threats, Protection and Attribution of Cyber Attacks on Critical Infrastructures}
\title{Threats, Countermeasures and Attribution of Cyber Attacks on Critical Infrastructures}

\author{Leandros Maglaras\affil{1}\affil{2}\fnoteref{1}, Mohamed Amine Ferrag\affil{3}\affil{4}, Abdelouahid Derhab\affil{5}, Mithun Mukherjee\affil{6}, Helge Janicke\affil{1}, Stylianos Rallis\affil{2}}

\address{\affilnum{1}Cyber Security Center, De Montfort University, Leicester, LE1 9BH, UK\\
\affilnum{2}General Secretariat of Digital Policy, Ministry of Digital Policy Telecommunications and Media, Athens, Greece \\
\affilnum{3}LabSTIC Laboratory, Guelma University, 24000 Guelma, Algeria\\
\affilnum{4}LRS Laboratory, Badji Mokhtar-Annaba University, 23000 Annaba, Algeria\\
\affilnum{5}Center of Excellence in Information Assurance (CoEIA), King Saud University, Saudi Arabia\\
\affilnum{6}Guangdong University of Petrochemical Technology, Maoming 525000, China\\
}

\abstract{
As Critical National Infrastructures are becoming more vulnerable to  cyber attacks, their protection becomes a significant issue for any organization  as well as a nation. Moreover, the ability to attribute is a vital element of avoiding impunity in cyberspace. In this article, we present main threats to critical infrastructures along with protective measures that one nation can take, and which are classified according to legal, technical, organizational, capacity building, and cooperation aspects. Finally we provide an overview of current methods and practices regarding cyber attribution and cyber peace keeping}


\fnotetext[1]{Corresponding author.  Email: \email{leandros.maglaras@dmu.ac.uk}}

\maketitle            

\section{Introduction}

Cyber security is currently one of the main concerns for Supervisory Control and Data Acquisition (SCADA) and Industrial Control Systems (ICS) operators. In fact, SCADA systems collect the data and monitor the automation processes, which are visualized to the operators of the system via human-to-machine interfaces. The operators can take control of the system remotely and issue commands such as opening a valve, setting a temperature point or starting/stopping a pump~\cite{maglaras2018cyber}. 
Recently, several countries have witnessed the impact of cyber threats to the critical infrastructures \cite{bristol}. For example, in December 2015, Ukraine was hit by massive power outage due to an outcome of SCADA cyber attack~\cite{UkraineAttack}. This caused about 230K people out of power for several hours. Another attack that was reported in 2016, although happened in 2013, targeted a small dam in Rye Brook, New York~\cite{DamAttack}. Based on a joint report by FBI and homeland security, the Wolf Creek Nuclear Operating Cooperation was  targeted~\cite{WolFCreek} in 2017. Although the nature of the attack was unknown, the impact could go beyond any nation. Recently, UK's general communications head quarters (GCHQ) and National Cyber Security Centers (NCSC) are concerned about suspicious attacks on UK energy sectors~\cite{UKPowerAttack}. The above are some of the examples, however, these cyber attacks can be related to any critical infrastructure, such as oil and gas industry, traffic signal, water sewage building, transportation, and digital infrastructure.

It is often observed that critical national infrastructures (CNI) are controlled, even in part, by private-sector companies. Therefore, cyber defence of any nation has to play a significant role in privately operated networks for the CNI.  Many SCADA applications are nowadays using common operating systems such as Windows as well as well-known and vulnerable protocols like Transport Control Protocol (TCP). The security vulnerabilities are publicly available and famous events like  Black Hat are now discussing more about industrial systems, proving that hackers are also focusing on these systems \cite{knapp2014industrial}.  Moreover, CNIs continue to suffer information security incidents and breaches as a result of human errors even though humans are recognised as the weakest link with regard to information security \cite{EVANS2018}.

{\it Lifecycle of Cybersecurity:}
Similar to other information technology (IT) processes, cyber security often follows a lifecycle model of prediction, protection, detection, and reaction.  The details are discussed as follows:

\begin{enumerate}
	\item During the {\it prediction phase,} each organization (or nation) needs to consider all proactive measures to identify potential attackers along with their intentions and the methods that they are going to use. This step can be implemented by collecting cyber threat intelligence and conducting risk management~\cite{ralston2007cyber}. Defining  scope and objectives for Risk Management, the external and internal environment of CNIs, conducting a risk assessment and producing appropriate risk mitigation plans are all steps that need to be taken, following a standardized methodology, e.g. ISO 31000. 

\item During protection phase, the organization applies hardware and software measures that are needed in order to accomplish its security goals, following the results of the risk assessment phase. 

\item During the detection phase, the organization needs to implement monitoring mechanisms along with intrusion detections systems \cite{maglaras2014intrusion},  which can distinguish legitimate from abnormal behavior or normal from malicious network traffic inside the system. 

\item The last phase of cyber security lifecycle includes all the  processes and methods that the organization (or nation) needs to have in place for incident notification and management, along with appropriate mitigation, recovery, and business continuity  plans. In the core of the cyber security lifecycle lies the cyber threat intelligence, which is the process of collecting data and deriving meaningful information for the system. 

\end{enumerate}

SCADA systems are nowadays the  targets  of cyber attackers, and it is worthwhile to note that attacking them affects a substantial number of persons, potentially causing significant damage and ultimately threatening human lives \cite{cook2016measuring}. Post-event investigation has frequently linked these attacks to the exploitation of vulnerabilities deeply rooted in the ICS design philosophy which focuses on availability rather than security.

In this article, we summarize the main issues in regulations for cyber security and outline several aspects of policy making to tackle cyber attacks on the CNIs. The rest of the paper is organized as follows. Section~2 discusses the main threats that critical infrastructures (CI)  are facing today. Section~3 presents several measures for the protection of CI. Techniques for cyber attack attribution are discussed in Section~4. Finally, conclusions are drawn in Section~5.

\section{Threats to Critical Infrastructures}

\begin{figure}[h]
\centering
\includegraphics[width=0.7\columnwidth]{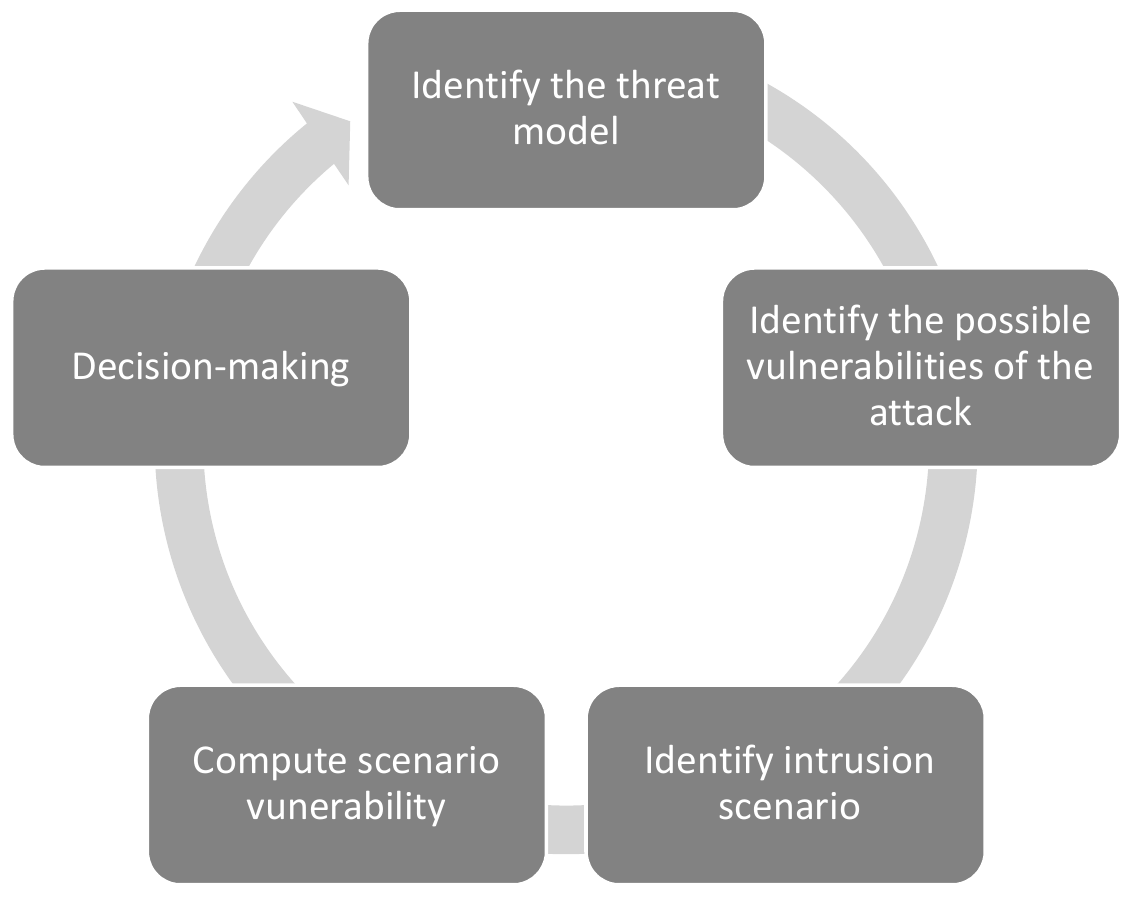} 
\caption{Procedures of vulnerability assessment of cybersecurity for critical infrastructures}
\label{fig:Fig1}
\end{figure}

Vulnerability assessment of cybersecurity is identified by five main phases, namely, 1) Identify the threat model, 2) Identify possible vulnerabilities of the attack, 3) Identify intrusion scenarios , 4) Compute scenario vulnerability, and 5) Decision-making, as presented in Fig.~\ref{fig:Fig1}.

In order to evaluate vulnerability indices on cybersecurity of critical infrastructures, Ten et al. \cite{F4} proposed two main procedures, namely, 1) cybersecurity conditions and 2) evaluation of vulnerability indices. The cybersecurity condition assessment is measured by a number $X$, which assumes the value of 0.33, 0.67, or 1. A low value indicates that the system condition is invulnerable, while the value 1 indicates that the system is vulnerable.  For the second procedure, the authors proposed four steps to assess the security vulnerability, namely, 1) identifying the intrusion scenarios; 2) evaluating vulnerability indices for the system, intrusion scenarios, and attack leaves; 3) port auditing; and 4) password strength evaluation.

\begin{figure}[h]
\centering
\includegraphics[width=0.9\columnwidth]{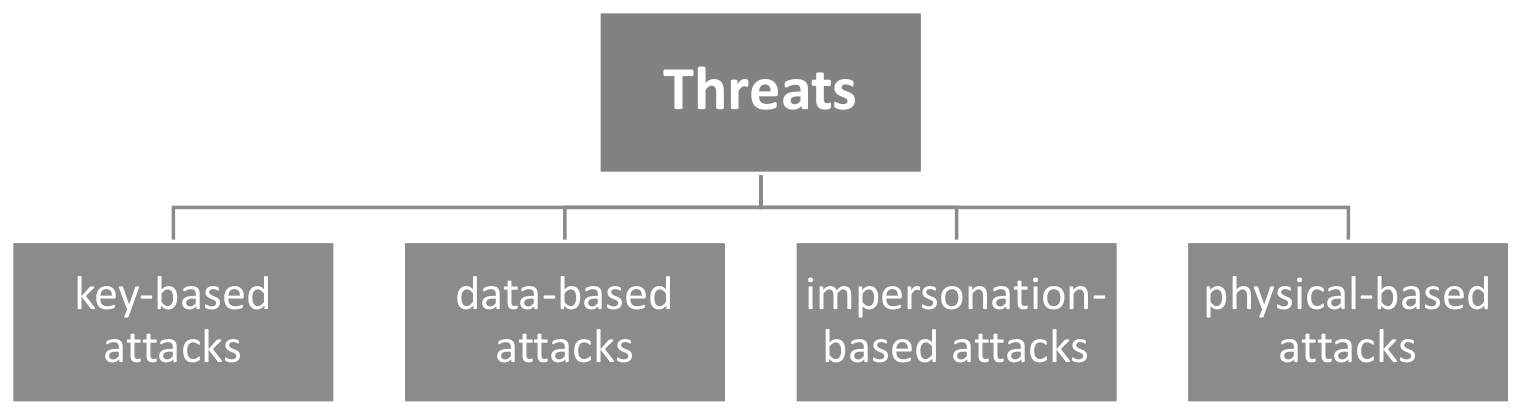} 
\caption{Classification of threats for SCADA system in Smart Grids}
\label{fig:Fig2}
\end{figure}

Since we are moving to the era of IoT, there are two categories of attacks:  

\begin{itemize}
\item The attacks on back-end IoT devices. This category of attacks is connected directly with critical infrastructures. Under this category, we can find the following attacks: attacks on SCADA systems and attacks analysis for critical infrastructure.

\item The attacks based on end-user IoT devices. This category of attacks is not connected directly with critical infrastructures. Under this category, we can find the following attacks: IoT-enabled attack vectors, IoT-enabled attacks on healthcare critical infrastructure, IoT-enabled attacks on intelligent transportation systems, IoT-enabled attacks on 5G cellular infrastructure.
\end{itemize}

\subsection{Attacks on SCADA systems}
SCADA systems and substations are now interconnected with other systems thanks to the {P}ower {S}ystem {C}ommunication (PSC) systems~\cite{F5}. SCADA system is the core of smart grid decision making. This interconnection between SCADA system and smart grid creates new possibilities and threats. Classification of threats in smart grids is done using different criteria such as passive or active, internal or external, etc.

In \cite{F6},  authors classified threats in smart grids into four categories, including, (1) key-based attacks, (2) data-based attacks, (3) impersonation-based attacks, and (4) physical-based attacks, as presented in Fig. \ref{fig:Fig2}. In order to detect attacks against critical infrastructures, Zonouz et al. \cite{F7} proposed a cyber-physical security state estimation framework, named SCPSE, which estimates the cyber-physical security state of a power grid. The SCPSE framework uses stochastic information fusion algorithms and merges sensor information from both cyber and electrical infrastructures. By using information provided by alerts from intrusion detection systems, the SCPSE framework can identify malicious measurement corruptions.

\subsection{Attack analysis for critical infrastructure}
Several techniques use the attack tree model to analyze the attacks targeting the critical infrastructures, as discussed by Fujita et al. \cite{F8}. Specifically, the authors proposed a systemic integration of granular computing and resilience analysis for critical infrastructures. This resilience analysis uses three tools, namely, 1) Three-way decisions as a tool for cognitive analysis, 2) Granular structures based on binary relations, 3) An approach based on the hierarchical granular modeling; and 4) Dominance-based rough sets as a tool to understand what are the parts of a critical infrastructure that are not performing well.

\subsection{IoT-enabled attack vectors}
Modeling of IoT-enabled attack vectors against critical infrastructures and services, as proposed by Stellios et al. \cite{F9}, contains three main entities, namely, 1) adversary, 2) IoT device, and 3) actual target. 

\begin{itemize}
\item The adversary is characterized by the following three capabilities: required access to the IoT, technical skills, and required motivation.
\item IoT vulnerabilities are categorized by embedded vulnerabilities and network vulnerabilities.
\item The connectivity between the IoT device and the actual target is categorized by the following two types: Direct connectivity with a critical system, and indirect connectivity with a critical system.
\end{itemize}

\subsection{IoT-enabled attacks on healthcare critical infrastructure}
According to Gope and Hwang \cite{F10}, IoT-enabled attacks on healthcare critical infrastructure can be classified into three types of attack assumptions, namely, 1) Computational capabilities, 2) Listening capabilities, and 3) Broadcasting capabilities.

\begin{itemize}
\item  With computational capabilities, an adversary can launch data modification and impersonation attacks.
\item  By using listening capabilities, an attacker can perform eavesdropping and tracking of healthcare critical infrastructure.
\item  Based on broadcasting capabilities, an attacker can replay critical data of a healthcare critical infrastructure.
\end{itemize}

\subsection{IoT-enabled attacks on intelligent transportation systems}
For modeling IoT-enabled attacks on intelligent transportation systems, Petit and Shladover \cite{F11}categorized attackers in an automated vehicle system as, 1) Internal versus external, 2) Malicious versus rational, 3) Active versus passive, 4) Local versus extended, and 5) Intentional versus unintentional. 
For future autonomous automated vehicles, the following attack surfaces can be considered:

\begin{itemize}
\item  Infrastructure sign
\item  Machine vision
\item  Global positioning system
\item  In-vehicle devices
\item  Acoustic sensors
\item  Radar
\item  Light detection and ranging
\item  Material/structure on which the vehicles drive
\item  In-vehicle sensors
\item  Odometric sensors
\item  Electronic devices
\item  Maps for longitudinal and lateral directions
\end{itemize}

\begin{figure}[h]
\centering
\includegraphics[width=0.9\columnwidth]{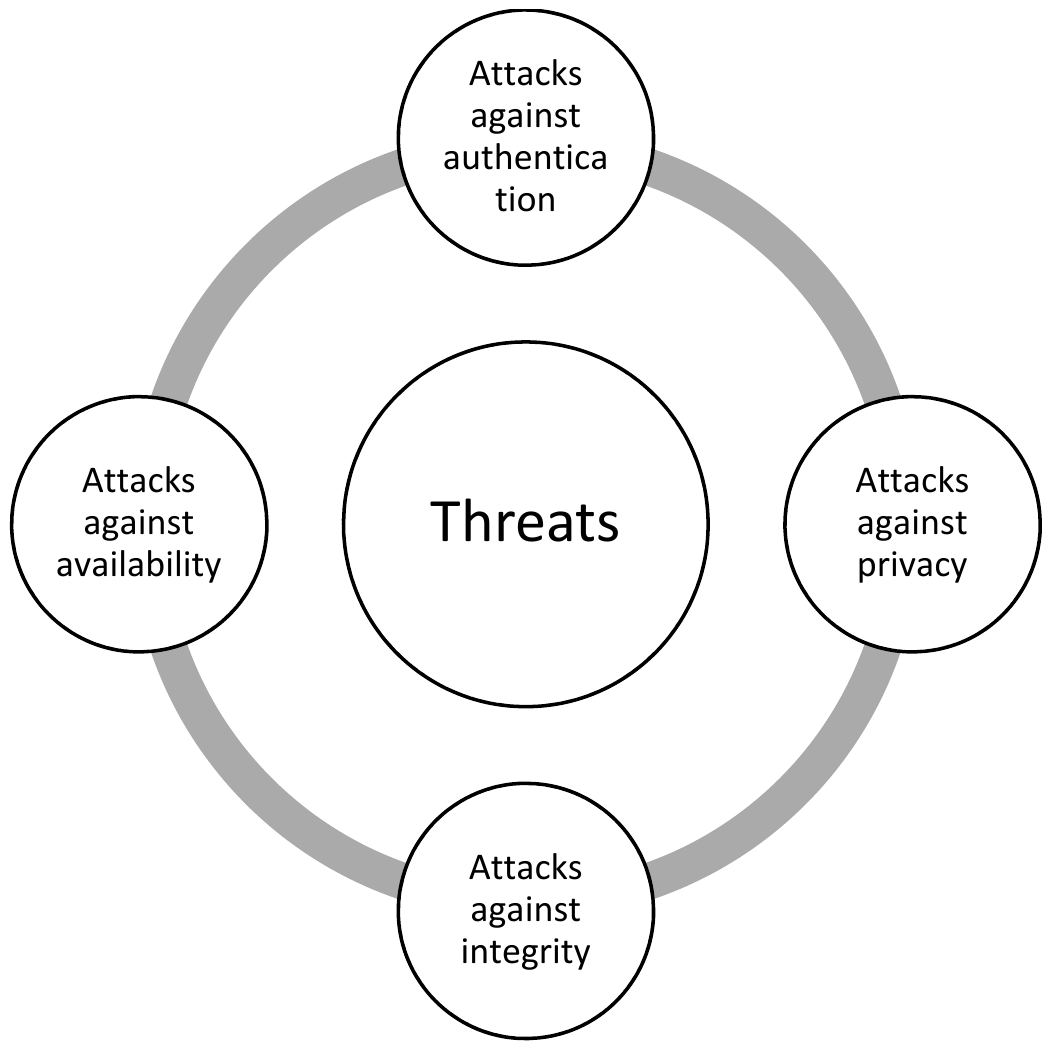} 
\caption{Classification of threats for 5G cellular infrastructure}
\label{fig:Fig3}
\end{figure}

\subsection{IoT-enabled attacks on 5G cellular infrastructure}
In general, a central base station cannot determine the behavior of the users in 5G cellular infrastructure. According to Geraci et al. \cite{F12}, the confidential messages in 5G cellular infrastructure can be eavesdropped by both: (i) users in the same cell and (ii) users in other cells.

In \cite{F13},  authors classified IoT-enabled attacks on 5G cellular infrastructure into four categories, including, (1) attacks against privacy, (2) attacks against integrity, (3) attacks against availability, and (4) attacks against authentication, as presented in Fig \ref{fig:Fig3}. Ahmad et al. \cite{F14,F15} in another interesting article provided an overview of different types of IoT-enabled attacks on 5G cellular infrastructure according to target point/network element, which can be SDN controller-switch communication, subscriber location, virtual resources, hypervisor, shared cloud resources, open air interfaces, unencrypted channels...etc.

\section{Protection of Critical Information Assets}
In this section, we propose some measures for the protection of Critical Infrastructure. The measures are divided into two main categories: \textit{cyber security measures} and \textit{cyber threat intelligence}. 

\begin{figure}[h]
\centering
\includegraphics[width=1.0\columnwidth]{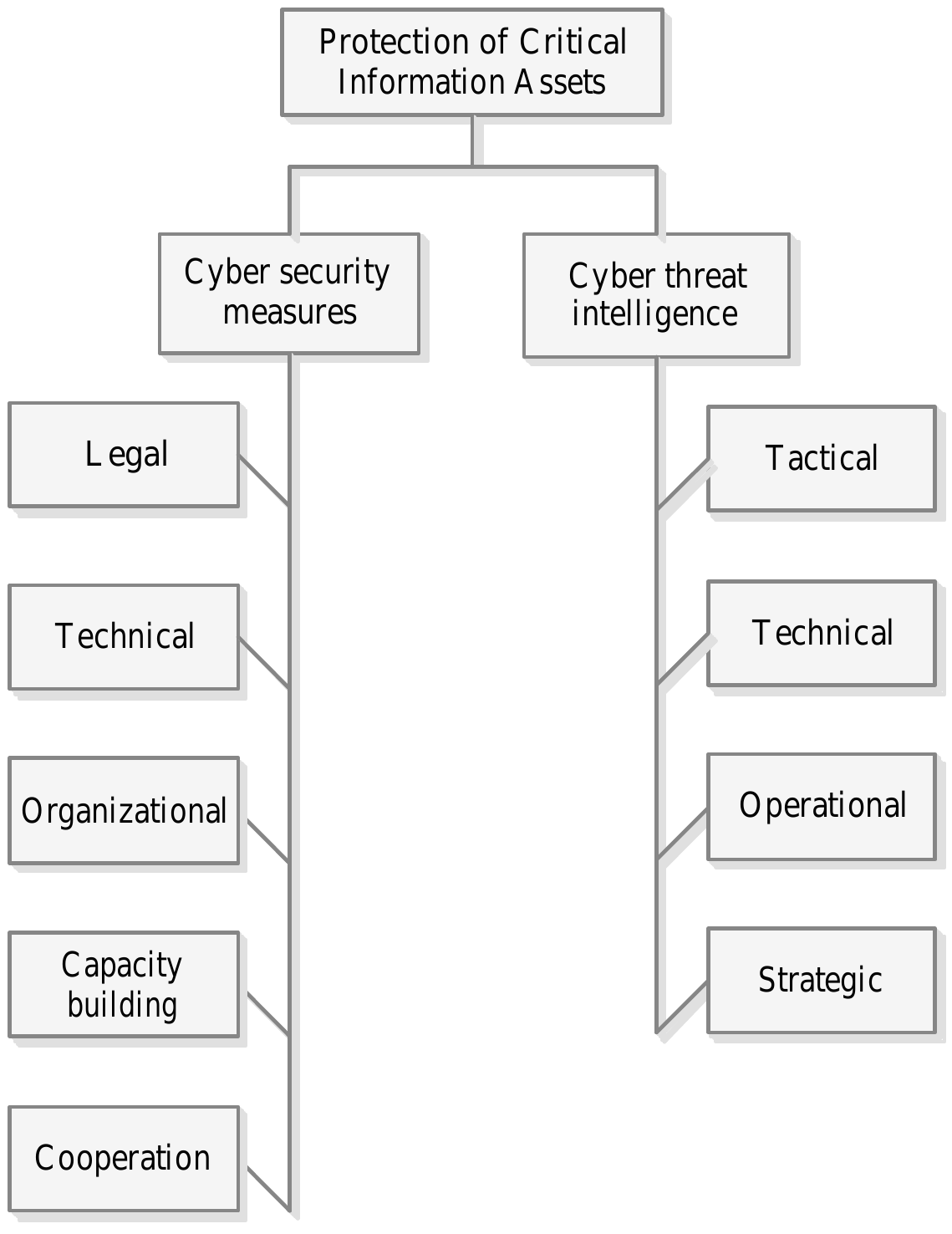} 
\caption{Measures for the protection of critical information assets}
\label{fig:Fig3}
\end{figure}

\subsection{Cyber security measures}

They are classified with respect to: legal, technical, organizational, capacity building, and cooperation aspects, as defined by the International Telecommunication Union (ITU) \cite{brahima2017global,pena2015global}.

\begin{itemize}

\item \textbf{Legal measures} aim to provide legislations and an implementable regulatory framework to protect the cyber space. As for CI, the following good practices can be recommended:

\begin{itemize} 

\item Ensure mandatory periodic assessment of CIs through information security audits

\item Check the compliance of software and hardware tools, which are used in the CI, with recognized security standards such as: ISO 27001.

\end{itemize}

\item \textbf{Technical measures} consider the technological tools (software and hardware) to prevent, detect,  mitigate, and respond to cyber attacks, such as:

\begin{itemize}
\item Implementation of internationally recognized security standards within organizations, especially the critical infrastructure ones.

\item Implementation of preventive and detective security tools such as: firewalls, Intrusion Detection System (IDS), Intrusion Prevention System (IPS), Antivirus/ Anti-malware.

\item Implementation of measures for physical security, access control, patching and upgrading, and forensics.

\item Development of an incident response capability.
\end{itemize}

\item \textbf{Organizational measures} are important for the proper implementation of any type of national initiative or policy. Under this aspect, we recommend the following:

\begin{itemize}
\item Develop a national critical infrastructure protection policy.
\item Develop a national framework for the implementation, evaluation, and maintenance of the cyber security policies.
\item Define an information security program for organizations.
\item Develop a national contingency plan.
\item Identify a national agency for the implementation of the critical infrastructure protection policy.
\item Conduct a national exercise to assess the cyber resilience of the CI.
\item Conduct security audits by organizations to check their cyber security preparedness.
\end{itemize}

\item \textbf{Capacity building measures} aim to enhance knowledge and know-how in order to promote cyber security. Under this aspect, we recommend the following:

\begin{itemize}
\item Encourage IT specialists in CI sectors to be certified under internationally recognized cyber security programs.
\item Conduct periodic awareness and training programs for employees.  
\end{itemize}

\item \textbf{Cooperation measures} aim to establish partnership between different stakeholders to increase cyber resilience of the organizations against cyber threats. Under this aspect, we recommend the following:

\begin{itemize}
\item Establish trusted information sharing mechanisms on threats and vulnerabilities between private and public stakeholders 
\item Establish a cooperation framework between industry and research to promote cyber security and increase resiliency against cyber attacks. 
\item Build a cooperation framework between countries on different aspects related to cyber security.
\item Contribute in international efforts to protect the cyber space
\end{itemize}

\end{itemize}

\subsection{Cyber threat intelligence}

Cyber threat intelligence refers to the collection of inteligence before a cyber attacker targets a victim system. The purpose is to help organizations understand and mitigate the risks related to zero-day exploits, Advanced Persistent Threats (APTs), internal and external threat actors. This allows organizations to adopt a proactive cybersecurity approach, and take preventive countermeasures in advance. Inteligence can be gathered from different sources such as: open source intelligence (OSINT), social media intelligence (SOCMINT), human Intelligence (HUMINT), technical intelligence, and intelligence from the deep and dark web. The United Kingdom's National Cyber Security Centre (NCSC) classifies threat intelligence into the following four classes \cite{chismon2015threat}:

\begin{itemize}
 \item 
 \textbf{Tactical Cyber threat intelligence}: This data is obtained from real-time monitoring of systems. It refers to real-time events and information related to adversary's actions inside the organization. Tactical threat intelligence is consumed by defenders to ensure that their incident response systems and investigations are prepared for the tactics. 
 
\item \textbf{Technical Cyber threat intelligence}: This data is consumed through technical means, e.g., suspected malicious IP address. It has a short lifetime as an adversary can for example change the IP address. Technical threat intelligence generally helps the the defenders to take preventive actions, e.g., blocking the suspected IP address.

\item \textbf{Operational Cyber threat intelligence}: This data gives details about a specific incoming attack such as: campaigns, malware, or cyber-weapon tools. It gives insignts that can guide and support the response to specific incidents, and help assessing the ability of the organization in determining future cyber threats·

\item \textbf{Strategic Cyber threat intelligence}: This data represents high-level information and a timely warning of cyber threats, which is consumed at the board level or by other senior decision-makers. Strategic cyber threat intelligence forms an overall picture of the intention and capabilities of malicious cyber actors and their impact on high-level business decisions.

\end{itemize}

\section{Attribution of Cyber Attacks}

When a cyber attack is launched against a CI, it is likely that some real-world physical revelation will follow \cite{maglaras2018cyber}. In some instances this could even lead to physical damage, injury, environmental effects or even loss of life. According to international or national law \cite{maglaras2018nis}, legal or regulatory investigation may be required, increasing the importance of attribution artefacts.  According to NIS directive, each Member State shall designate one or more national competent authorities for the security of network and information systems that can take the leading role in securing CIs. The responsible authority of the country will have to identify whether the incident was caused by an error in the operations, maybe a fault component, or whether the processes or devices were maliciously manipulated. Artefacts should be collected and kept in a way that both authenticity, integrity and usability are guaranteed.  

Researchers have surveyed individual technical approaches to attribution, including; traceback - where the traffic from a target device is recursively steppedback through its routing path to its originating source, honeypots - where vulnerable software and services are hosted in order to allow activities to be monitored among others.  

{\bf Traceback} is a class of methods that encompasses techniques by which the traffic from a target device is recursively stepped-back through its routing path to its originating source device \cite{kuznetsov2002evaluation}. Traceback can be supported by manual methods of traffic tracing or logging techniques supported from network devices. There is a third category of traceback that includes various methods of {P}robabilistic {P}acket {M}arking (PPM) \cite{song2001advanced}, and ICMP traceback (iTrace) \cite{bellovin2003icmp}.

{\bf Honeypots} approach the issue of attribution of attacks differently to Traceback methods. A honeypot is a system, or set of systems, where vulnerable software and services are hosted in order to allow activities to be monitored and logged. Several researchers have proposed the use of honeypots to protect important assets of critical infrastructure \cite{simoes2013use}. However, most of these honeypots are static systems that wait for the attackers. In order to increase the efficiency of honeypots they need to be as realistic as possible. In \cite{lin2017generating} authors introduce a honeypot network traffic generator that mimics a genuine control system in operation.

{\bf Digital Forensics} is a broad subject which involves the recovery, acquisition and investigation of digital evidence. One technique that could be used is live forensics where data acquisition takes place while the system is operational. In order to avoid system crash, especially for the SCADA systems that exist in the core of each CI, authors in \cite{ahmed2012scada} propose the use of fail over systems. In either case post incident investigator will compete with recovery efforts, which will most likely destroy evidence.

{\bf Network Forensics} field primarily involves two stages: collecting network messages and analyzing network messages. An organization must identify points in the network where they wish to collect network data. Again special care should be given regarding SCADA operation requirements \cite{mahmood2010network}.

{\bf Malware Analysis} can be split into two areas:
behavioural analysis and code analysis. Behavioral analysis examines the way that malware interacts with the environment \cite{shukla2008application} while code analysis examines the code that makes up the malware \cite{bayer2006dynamic}.

Measuring the performance of attribution attacks is an open issue although several methods have been proposed \cite{nicholson2012taxonomy}. The development of modelling strategies for evaluating cyber attacks \cite{pipyros2018new} are also important. Cook et al. \cite{cook2016attribution} have used six individual metrics, as summarized in Table $1$, to measure the effectiveness of each attribution in the context of ICS that can be applied to CIs. 

\begin{table*}
\begin{center}
\begin{tabular}{ |r|l|}
\hline
 \bf{Metric} & \bf{Description}\\\hline
  Performance &    Ability to provide attribution functions without degrading CI performance\\
Reliability  &   Ability to provide attribution without  affecting  operating and safety processes\\
 Extent  &   Ability to monitor traffic to provide a full picture of network behaviour \\
Coherence  & Cross-reference traffic with device behaviours to permit inspection of command execution\\
Identification  & Ability to identify the attacker from behaviours or technical signatures\\
Intent  & Ability to determine the purpose of the attack in order to support a prosecution \\
\hline
\end{tabular}
\end{center}
\label{tab-1}
\caption{Metrics of attribution} 
\end{table*}

Attacks may sometimes originate from another nation and attribution becomes a question of which country or law enforcement agency has the responsibility and authority to investigate, under which legal framework the perpetrators can be prosecuted, and which laws apply.
This transnational issue was analyzed in the Tallinn Manual on the International Law Applicable to Cyber Warfare \cite{schmitt2013tallinn}. In the same context authors in \cite{robinson2018introduction}  argued that as cyber warfare becomes an increasing part of wider conflict, peacekeeping organizations such as the United Nations will probably need to perform cyber peacekeeping.

In order to perform cyber peacekeping, UN or the competent international body, should also consider peace enforcement in order to protect civilians. While such an event seems valuable, the feasibility is be questionable.  Research into cyber warfare~\cite{robinson2015cyber} has shown that there is still no answer to questions such as what constitutes an armed attack in the cyber space or what the ethical boundaries of cyber warfare are. In the cyber domain, it is difficult to foresee the Security Council of th UN agreeing to enforce peace based upon a cyber conflict \cite{robinson2018developing}.  In depth study of how cyber peace enforcement could work and the value it could bring would be useful.

A big challenge is the technical challenges CNI presents. Facilities such as power plants and water facilities have properties, which make observation and monitoring more challenging than standard network monitoring techniques~\cite{nicholson2012taxonomy}. The use of proprietary protocols, air-gapping and a 24/7 availability requirement means that monitoring of events on these systems will require a specialised set of skills that are in high demand globally.  A discussion on how the right expertise can be secured in the necessary numbers, and at a price that is within an operation's budget is a necessary future research topic.

Multi-lateral cooperation can be used in order to detect attacks and perform attribution, because even the largest intelligence organizations have limited human, technological and budgetary capacities to achieve global coverage \cite{Saalbach2017}. US has established already after World War II the declassified 5-eyes cooperation with UK, Canada, Australia and New Zealand and in response to 9/11 a wider cooperation the 9-eyes cooperation including Denmark, France, Netherlands and Norway and finally the 14-eyes cooperation additionally including Belgium, Italy, Spain, Sweden and Germany. 

\section{Conclusions}
As CNIs are vulnerable to cyber attacks, their protection  becomes a significant issue for any nation as well as an organization. In this article, we have summarized the primary attributes of cyber attacks to the critical infrastructures. We have further provided the protective cyber security measures that one nation can take, and which are classified according to legal, technical, organizational, capacity building, and cooperation aspects. Cyber threat intelligence is also an important protection aspect as it helps taking countermeasures in advance, and enables developing a proactive and predictive cyber security posture. Attribution of cyber attacks, especially when the latter originate from another nation, is questionable regarding which country or law enforcement agency has the authority to investigate and prosecute the penetrators and cyber peacekeeping is foreseen to become a reality.

\bibliographystyle{unsrt}
\bibliography{maglaras,ferrag}
\end{document}